\documentclass[12pt,letterpaper]{article}

\parskip=\medskipamount            % space between paragraphs (LaTeX)
\def\7#1#2{\mathop{\null#2}\limits^{#1}}        % puts #1 atop #2

\def\beee{\begin{equation}}
\def\eeee{\end{equation}}
\def\dggg{^{\dagger}}
			
\begin{document}

\bibliographystyle{unsrt}
\begin{center}
\textbf{CPT VIOLATION IMPLIES\\ VIOLATION OF LORENTZ INVARIANCE}\\
[5mm]
O.W. Greenberg\footnote{email address, owgreen@physics.umd.edu.}\\
{\it Center for Theoretical Physics\\
Department of Physics \\
University of Maryland\\
College Park, MD~~20742-4111}\\
University of Maryland Preprint PP-02-027\\
~\\
Version to appear in Physical Review Letters
\end{center}

\begin{abstract}

An interacting theory that violates $CPT$ invariance 
necessarily violates Lorentz invariance.  On the other hand,  
$CPT$ invariance is not sufficient for out-of-cone Lorentz invariance. 
Theories that violate $CPT$ by having different particle and 
antiparticle masses must be nonlocal.

\end{abstract}

A quantum field theory is Lorentz covariant in cone\cite{cone} 
if vacuum matrix 
elements of unordered products of fields (Wightman functions) are
covariant.  (What we call ``functions'' are
distributions in a space of ``generalized functions.'')  We assume
in-cone Lorentz covariance (actually Poincar\'e covariance) in this
paper.  A quantum
field theory is covariant out of cone if vacuum matrix elements of 
time-ordered products ($\tau$ functions) are covariant.  To calculate
the $S$ matrix we need $\tau$ functions, or similar functions, such
as retarded or advanced products ($r$ functions or $a$ functions).
We require covariance of a quantum field theory both in and out of cone
as the condition for Lorentz invariance of the theory; thus both the
Wightman functions and the $\tau$ (or $r$ or $a$) functions must be
covariant for the theory to be Lorentz invariant.

Res Jost proved the fundamental theorem\cite{jos} that weak local 
commutativity
at Jost points is necessary and sufficient for $CPT$ symmetry. 
Jost points
are spacetime points in which all convex 
combinations of the successive
differences are spacelike; i.e., if the Wightman function is
\beee
W^{(n)}(x_1,x_2, ...,x_n)=\langle 0|\phi(x_1)\phi(x_2) 
\cdots \phi(x_n)|0\rangle,
\eeee
a Jost point is an ordered set of $\{x_i\}$ such that all sums
$\sum_i c_i(x_i-x_{i+1}),\\
c_i \geq 0,~~\sum_i c_i >0,$ are spacelike.  
Weak local commutativity states that
\beee
W^{(n)}(x_1,x_2, ...,x_n)=W^{(n)}(x_n,x_{n-1}, ...,x_1).
\eeee
Clearly local commutativity implies weak local commutativity.

We will show that violation of $CPT$-invariance in 
{\em any} Wightman function\cite{wig} implies 
noncovariance of the related $\tau$ (or $r$ or $a$) function, and thus 
implies a violation of Lorentz invariance of the theory. We give our 
explicit
discussion in terms of a scalar theory; however analogous arguments 
apply for any spin.

Alan Kosteleck\'y and collaborators have studied $CPT$- and 
Lorentz-violating theories systematically.  Among their papers 
are\cite{col,kos}.  The proceedings of a meeting on $CPT$ and Lorentz
symmetry are in\cite{proc}.  Kosteleck\'y, \textit{et al}, have emphasized the
difference between observer (sometimes called ``passive'' in theories
without Lorentz violation) Poincar\'e
transformations in which coordinates
are transformed according to $x \rightarrow \Lambda^{-1}(x-a)$, 
where $\Lambda$ is a Lorentz transformation and $a$ is a spacetime 
translation, and particle (sometimes called ``active'' in theories
without Lorentz violation) Poincar\'e 
transformations in which fields are transformed according to
\beee
U(a,\Lambda)\phi(x)U(a,\Lambda)\dggg=\phi(\Lambda x + a),
~U(a, \Lambda)|0\rangle=|0\rangle.
\eeee
The analogous distinction exists for $CPT$ symmetry. Unless stated
otherwise, we always consider observer symmetries in this paper.
If $CPT$ is violated for any $\tau$ function,
which implies that weak local commutativity is violated for that 
function, then the corresponding $\tau$ function 
is not Lorentz covariant
(not Lorentz invariant for the scalar case we discuss explicitly), 
and the theory is not Lorentz invariant.  A general $\tau$ 
function is given in terms of Wightman functions as
\beee
\tau^{(n)}(x_1, x_2, \cdots x_n)=
\sum_P\theta(x^0_{P_1}, x^0_{P_2}, ..., x^0_{P_n})
W^{(n)}(x_{P_1},x_{P_2}, ...,x_{P_n}),
\eeee
where $\theta$ enforces 
$x^0_{P_1}\geq x^0_{P_2}\geq...\geq x^0_{P_n}$.  In order for the
$\tau$ function to be Lorentz covariant, Lorentz transformations
that reverse the time order of points while leaving the Wightman
function invariant must not change the $\tau$ function.  This
requires the equality of Wightman functions of permuted field
orders when the relative distances are spacelike.  If the functions
are at a Jost point and the $\{x_i\}$ are such that all the successive
time differences are positive then there is an observer
Lorentz transformation that leaves the Wightman functions invariant, 
but
makes all the successive time differences negative.  Invariance of the
$\tau$ function requires that the original Wightman function 
and the one with the fields in completely reversed order have
the same value.  This is precisely the condition of weak local 
commutativity which is necessary and sufficient for $CPT$ invariance
of the corresponding matrix element.  Thus if $CPT$ invariance does not
hold for this matrix element, then the $\tau$ function is not 
Lorentz invariant and the theory is not Lorentz invariant.  This
argument does not apply to a noninteracting theory for which $\tau$
functions need not be considered. Thus we
have demonstrated the main result of this paper, \emph{If $CPT$
invariance is violated in an interacting quantum field theory, then 
that theory also violates Lorentz invariance.}  

%Here
%we disagree with the assertion\cite{bar, bar2} that their model 
%that violates $CPT$ can be Lorentz invariant. 

On the other hand, weak local commutativity insures
only one of the equalities among Wightman functions of fields in
permuted orders that are necessary for out-of-cone
Lorentz covariance of the $\tau$
functions, so $CPT$ invariance is not sufficient for out-of-cone
Lorentz invariance.

We remark that weak local commutativity of Wightman functions implies
weak local commutativity of truncated Wightman functions.  We see no 
reason why violations of weak local commutativity cannot occur 
independently
in truncated Wightman functions of different order. Even if the 
two-point
Wightman function obeys weak local commutativity (which is the same
condition as local commutativity for this case) which implies equal 
masses
for the particle and antiparticle, weak local commutativity can be 
violated
for the higher Wightman functions, and thus $CPT$ can be violated in 
scattering and other physical processes even when the masses of 
particle and antiparticle are equal.

These results are relevant to theories in which the effective 
four-dimensional theory comes from a higher dimensional 
theory: if the effective four-dimensional theory violates $CPT$
symmetry it also violates Lorentz invariance.

We consider the case in which the particle and 
antiparticle have different masses\cite{bar,bar2,mur} specifically.  
We disagree with the assertion\cite{bar, bar2} that their model 
that violates $CPT$ can be Lorentz covariant. 
We discuss the case of a charged scalar field; the results
for other spin fields are qualitatively the same.  We use covariant
normalization for the annihilation and creation operators.
We take the (Bose) commutation relations for the asymptotic
(in or out) particles and antiparticles (we drop the labels in or out 
to simplify notation) for any observer to be
\beee
[a(\mathbf{p}),a\dggg(\mathbf{p}^{\prime})] = 2E(\mathbf{p}) 
\delta(\mathbf{p}-\mathbf{p}^{\prime}),
~E(\mathbf{p})= \sqrt{\mathbf{p}^2+m^2},            \label{a}
\eeee
\beee
[b(\mathbf{p}),b\dggg(\mathbf{p}^{\prime})] = 2\bar{E}(\mathbf{p}) 
\delta(\mathbf{p}-\mathbf{p}^{\prime}),
~\bar{E}(\mathbf{p})= \sqrt{\mathbf{p}^2+\bar{m}^2}.    \label{b}
\eeee
For $m^2 \neq \bar{m}^2$ these commutation relations violate $C$ and 
$CPT$, but not necessarily $P$ and $T$.  These relations do not seem 
to occur in the general analysis of Kosteleck\'y, \textit{et 
al}\cite{col,kos,proc};  
we believe this is because Kosteleck\'y, \textit{et al}, assume observer Lorentz
covariance both in and out of cone, while we only assume in-cone 
observer Lorentz covariance.

We choose the Hamiltonian for this case to be
\beee
H=\int[\frac{d^3p}{2E(\mathbf{p})}E(\mathbf{p}) 
a\dggg(\mathbf{p})a(\mathbf{p})
+\frac{d^3p}{2\bar{E}(\mathbf{p})}\bar{E}(\mathbf{p})
b\dggg(\mathbf{p})b(\mathbf{p})].
\eeee
The usual commutation relations of the Lie algebra of the Poincar\'e
group,
\beee
[P^{\mu},~P^{\nu}]=0,~~[M^{\mu \nu},~P^{\lambda}]=
i(\eta^{\nu \lambda}P^{\mu}-\eta^{\mu \lambda}P^{\nu}),
\eeee
\beee
[M^{\mu \nu},~M^{\alpha \beta}]=i(\eta^{\mu \beta} M^{\nu \alpha}
-\eta^{\mu \alpha} M^{\nu \beta}+\eta^{\nu \alpha} M^{\mu \beta}
-\eta^{\nu \beta} M^{\mu \alpha})
\eeee
are satisfied by the replacements
\beee
P^{\mu} \rightarrow p^{\mu},~~M^{\mu \nu} \rightarrow 
i(p^{\mu} \frac{\partial}{\partial p_{\nu}}-
p^{\nu} \frac{\partial}{\partial p_{\mu}}).
\eeee
Carrying this over to the field operators gives Hermitian operators for
the generators of the Lie algebra,
\beee
P^{\mu}=\int[\frac{d^3p}{2E(\mathbf{p})}
a\dggg(\mathbf{p})p^{\mu}a(\mathbf{p})+
\frac{d^3p}{2\bar{E}(\mathbf{p})}
b\dggg(\mathbf{p})\bar{p}^{\mu}b(\mathbf{p})],
\eeee
\beee
M^{\mu \nu}=\int[\frac{d^3p}{2E(\mathbf{p})}a\dggg(\mathbf{p})
i(p^{\mu} \frac{\partial}{\partial p_{\nu}}-
p^{\nu} \frac{\partial}{\partial p_{\mu}})a(\mathbf{p})
+\frac{d^3p}{2\bar{E}(\mathbf{p})}
b\dggg(\mathbf{p})i(\bar{p}^{\mu} \frac{\partial}{\partial \bar{p}_{\nu}}-
\bar{p}^{\nu} \frac{\partial}{\partial \bar{p}_{\mu}})b(\mathbf{p})].
\eeee
Thus the free fields carry a representation of the Poincar\'e
algebra.  In particular, the Hamiltonian, $H=P^0$, generates time 
translations. As we will show below, the problem with Lorentz 
invariance only occurs out of cone 
and shows up explicitly when there are interactions.

We construct the spacetime dependence of the fields by using the
generators of translations,
\beee
e^{i P \cdot x}a(\mathbf{p})e^{-i P \cdot x}=e^{-i p \cdot x} a(\mathbf{p})
\eeee
Thus the $x$-space fields are
\beee
\phi(x)=\frac{1}{(2 \pi)^{3/2}}\int 
[\frac{d^3p}{2 E(\mathbf{p})}a(\mathbf{p})
e^{-i p \cdot  x}+
\frac{d^3p}{2 \bar{E}(\mathbf{p})}b\dggg(\mathbf{p})
e^{i \bar{p} \cdot x}],
\eeee
\beee
\phi\dggg(x)=\frac{1}{(2 \pi)^{3/2}}
\int [\frac{d^3p}{2 \bar{E}(\mathbf{p})}b(\mathbf{p})
e^{-i \bar{p} \cdot x}+
\frac{d^3p}{2 E(\mathbf{p})}a\dggg(\mathbf{p})e^{i p \cdot  x}],
\eeee
where $p^0=E(\mathbf{p})$ for terms with $a$ and $a\dggg$, and
$\bar{p}^0=\bar{E}(\mathbf{p})$ for terms with $b$ and $b\dggg$.

In this case the only truncated vacuum matrix elements are the 
two-point functions, which are covariant (invariant for this scalar 
case),
\beee
\langle 0|\phi(x) \phi\dggg(y)|0 \rangle =\Delta^{(+)}(x-y;m^2)
=\frac{1}{(2 \pi)^3}\int \frac{d^3p}{2E(\mathbf{p})}e^{-ip \cdot x},
\eeee
\beee
\langle 0|\phi\dggg(x) \phi(y)|0 \rangle =\Delta^{(+)}(x-y;\bar{m}^2)
=\frac{1}{(2 \pi)^3}\int \frac{d^3p}{2\bar{E}(\mathbf{p})}
e^{-i\bar{p} \cdot x}
\eeee
The field is local in sense (iii) defined below if the commutator 
vanishes at spacelike separation,
\beee
[\phi(x), \phi\dggg(y)]= 0,~(x-y)^2 <0.
\eeee
For this to hold the vacuum matrix element, $[\Delta^{(+)}(r;m^2)
-\Delta^{(+)}(-r;\bar{m}^2)]$, $r^{\mu}=x^{\mu}-y^{\mu}$,
must vanish at for $r^2<0$. The asymptotic limit for 
$\sqrt{-r^2} \rightarrow \infty$ is
\beee
\langle 0|[\phi(x), \phi\dggg(y)]|0 \rangle \rightarrow (-r^2)^{-3/4}
(e^{-m \sqrt{-r^2}}-e^{-\bar{m} \sqrt{-r^2}});
\eeee
this requires $m^2=\bar{m}^2$.

Under Poincar\'e transformations the two-point $\tau$ function, which 
is the Feynman propagator,
\beee
\langle 0|T(\phi(x) \phi\dggg(y))|0 \rangle = 
\theta(x^0-y^0) \langle 0|\phi(x) \phi\dggg(y)|0 \rangle +
\theta(y^0-x^0) \langle 0|\phi\dggg(y) \phi(x)|0 \rangle
\eeee
becomes
\begin{eqnarray}
\langle 0|T(\phi(\Lambda^{-1}(x-a)) 
\phi\dggg(\Lambda^{-1}(y-a)))|0 \rangle &=&  
\theta((\Lambda^{-1}(x-y))^0) \langle 0|\phi(x) \phi\dggg(y)|0 \rangle +
     \nonumber                                       \\
& & \theta((\Lambda^{-1}(y-x))^0) \langle 0|\phi\dggg(y) \phi(x)|0 \rangle,
\end{eqnarray}
where we used the translation and Lorentz invariance properties of
$\Delta^{(+)}(x-y;m^2)$ and of
$\Delta^{(+)}(y-x;\bar{m}^2)$.
If $x-y$ is spacelike a Lorentz transformation can transform a 
vector with $x^0 > y^0$ into one with $y^0 > x^0$, which changes the 
value of the propagator from $\Delta^{(+)}(x-y;m^2)$ to 
$\Delta^{(+)}(y-x;\bar{m}^2)$.  Thus the propagator 
is not covariant unless the vacuum matrix element of the commutator,
$\Delta^{(+)}(x-y;m^2)
-\Delta^{(+)}(y-x;\bar{m}^2)$, vanishes at spacelike separation and, as 
shown above, this
happens only if $m^2=\bar{m}^2$\cite{particle}.
 
Straightforward
calculation of the $\tau$ function in momentum space gives
\beee
[2i(E(\mathbf{p})-p^0-i\epsilon)E(\mathbf{p})]^{-1}
+[2i(\bar{E}(\mathbf{p})+p^0-i\epsilon)\bar{E}(\mathbf{p})]^{-1}. \label{prop}
\eeee
For $m^2=\bar{m}^2$ this reduces to the invariant form 
$i/(p^2-m^2+i\epsilon)$ as it should.

To illustrate the effect of the noninvariance of this propagator as 
viewed by different observers, assume
the propagator mediates a scalar $s$-channel process 
$k_1+k_2 \rightarrow k_1^{\prime}+k_2^{\prime}$, 
$k_1^2=k_1^{\prime~2}=m_1^2$, 
$k_2^2=k_2^{\prime~2}=m_2^2$.  If the propagator were Lorentz invariant 
an observer who saw the total momentum to be zero
(call this the center-of-mass frame) 
would find the same result as an observer who saw the momentum of one 
of the particles, say particle 2 to be zero (call this the lab frame).  
(Since the propagator in 
momentum space is just the Fourier transform of the propagator in 
position space, we know already that these results will not agree.  
The purpose of the following calculation is to show that in the 
high-energy limit the results disagree qualitatively.) In the 
center-of-mass frame, $\mathbf{p}=0$,
$s=(\sqrt{\mathbf{k}^2+m_1^2}+\sqrt{\mathbf{k}^2+m_2^2})^2$, 
$E(\mathbf{p})=m$, 
$\bar{E}(\mathbf{p})=\bar{m}$. The propagator is
\beee
propagator|_{cm}=\frac{m^2+\bar{m}^2+
(\bar{m}-m)\sqrt{s}}{2i m \bar{m}(m-\sqrt{s})(\bar{m}+\sqrt{s})}.
\eeee
In the lab frame, $\mathbf{p}=\mathbf{k_1}$,
$s= m_1^2+m_2^2+2m_2 \sqrt{k_1^2 + m_1^2}$, 
$E(\mathbf{p})=\sqrt{\mathbf{k}_1^2+m^2}$, $\bar{E}(\mathbf{p})=m_2$. The propagator
is given by Eq.(\ref{prop}), with 
\beee
E(\mathbf{p})=
\frac{\sqrt{s^2-2(m_1^2+m_2^2)s+(m_1^2-m_2^2)^2+4m_2^2m^2}}{2m_2},
\eeee
$\bar{E}(\mathbf{p})$ is similar, but with $m$ replaced by $\bar{m}$,
and $p^0=(s+m_2^2-m_1^2)/2m_2$.
In the limit
$s\rightarrow \infty$,
\beee
propagator|_{cm}\rightarrow \frac{i(\bar{m}-m)}{2m \bar{m}\sqrt{s}},
\eeee
\beee
propagator|_{lab}\rightarrow \frac{i}{s}.
\eeee
Thus the large-$s$ behaviors of the amplitude differ 
qualitatively in the two different frames.  If $m=\bar{m}$ both
propagators go to $i/s$ for large $s$.  For a resonant amplitude with
$s$ near $m^2$ or $\bar{m}^2$, this noninvariance of the propagator will
lead to noninvariance of the scattering cross section.

Next we discuss the question of nonlocality when the masses of the 
particle and antiparticle differ.

Nonlocal quantum field theories were discussed extensively in the 1950's
as a possible way to remove the ultraviolet divergences of quantum field
theories\cite{kri,blo,ebe,pei,mar}.  The nonlocality was always 
introduced in the interaction
terms, not in the quadratic terms that correspond to the free 
Hamiltonian.
Nonlocal theories failed as a mechanism to solve the divergence
problem.  Surprisingly, some of these authors derived conditions under 
which the acausality due to nonlocality is restricted.

The property of locality can have three different meanings for a quantum
field theory, (i) the fields enter
terms in the Hamiltonian and the Lagrangian at the same spacetime point, 
(ii) the
observables commute at spacelike separation, and (iii) the fields 
commute (for integer spin fields) or anticommute (for odd 
half-integer spin
fields) at spacelike separation.  Theories in which (i) fails can still
obey (ii) and (iii); for example, quantum electrodynamics in Coulomb gauge.
Theories in which (iii) fails can still obey (i) and (ii); for example,
parastatistics theories of order greater than one. 
We have already shown above that the theory
in which $CPT$ is
violated due to having different masses for the particles and antiparticles
is nonlocal in sense (iii).  We expect that such a theory will be nonlocal
in sense (ii), but we do not show this here. Next we do show that such a 
theory is nonlocal in sense (i).
  
We discuss the case of a charged scalar field explicitly; the results
for other spin fields are qualitatively the same.  We use the annihilation 
and creation operators given in Eq.(\ref{a},\ref{b}).
We calculate the free Hamiltonian
for such fields in two ways.  First, we calculate the Hamiltonian 
assuming only first derivatives of the fields enter and, secondly, we
do the calculation allowing higher derivatives.

With only first order derivatives, we find
\beee
a({\textbf p})=\frac{4E_p \bar{E}_p}{(E_p+\bar{E}_p)^2}
[\phi_p+\frac{E_p-\bar{E}_p}{2\bar{E}_p}
\phi_{(-\bar{E}_p,{\textbf p})}
exp(i(E_p+\bar{E}_p)x^0)],
\eeee                                                       \label{aa}
\beee
b(\bar{\textbf {p}})=-\frac{4E_p \bar{E}_p}{(E_p+\bar{E}_p)^2}
[\phi_{(-\bar{E}_p,{\textbf p})}-\frac{E_p-\bar{E}_p}{2E_p}\phi_p
exp(-i(E_p+\bar{E}_p)x^0)].
\eeee                                                       \label{bb}
Here we used the definition
\beee
\phi_p=\frac{i}{(2 \pi)^{3/2}}\int d^3x e^{i p \cdot x} 
\stackrel{\textstyle \leftrightarrow}{\partial_{x^0}} \phi(x).
\eeee                                                       \label{aaa}
The explicit nonlocal form of the Hamiltonian,
\beee
H=\frac{1}{2}\int d^3p[a\dggg(\mathbf{p})a(\mathbf{p})+
b\dggg(\bar{\mathbf{p}})b(\bar{\mathbf{p}})]
\eeee                                                       \label{bbb}
expressed in terms of $\phi(x)$ and $\phi\dggg(x)$ follows from
Eq.(27-29); it is nonlocal in space,
complicated, and not informative.

Allowing higher derivatives, however, 
leads to a relatively simple form for the
Hamiltonian with the $\Delta^{(+)}$ function
as the kernel that gives the nonlocality.  The result is
\begin{eqnarray}
H &=& \frac{2i}{(2\pi)^3 (m^2-\bar{m}^2)^2}
\{ \int d^3x d^3x^{\prime} \frac{\partial 
\Delta^{(+)}(x-x^{\prime}; m^2)}{\partial(x^0-x^{\prime~0})} \nonumber\\
& \times &
\stackrel{\textstyle \leftrightarrow}{\frac{\partial}{\partial x^0}}
\stackrel{\textstyle \leftrightarrow}
{\frac{\partial}{\partial x^{\prime~0}}}
(\partial_x \cdot \partial^x + \bar{m}^2)
(\partial_{x^{\prime}} \cdot \partial^{x^{\prime}} + \bar{m}^2)
\phi\dggg(x)\phi(x^{\prime}) \nonumber \\
&+& \int d^3x d^3x^{\prime} \frac{\partial 
\Delta^{(+)}(x-x^{\prime}; \bar{m}^2)}{\partial(x^0-x^{\prime~0})} \nonumber\\
&\times&
\stackrel{\textstyle \leftrightarrow}{\frac{\partial}{\partial x^0}} 
\stackrel{\textstyle \leftrightarrow}
{\frac{\partial}{\partial x^{\prime~0}}}
(\partial_x \cdot \partial^x + m^2)
(\partial_{x^{\prime}} \cdot \partial^{x^{\prime}} + m^2)
\phi(x)\phi\dggg(x^{\prime})\}.                     \label{h}
\end{eqnarray}
The apparent singularity in Eq.(\ref{h}) due to the factor 
$(m^2-\bar{m}^2)^{-2}$ is removed by the Klein-Gordon operators and
the Klein-Gordon scalar products in this equation.

To summarize, we have demonstrated that $CPT$ invariance is necessary, 
but not sufficient, for Lorentz invariance of an interacting quantum 
field theory. We noted that violations of $CPT$ can
occur independently in different truncated Wightman functions.

We also showed that if one explicitly chooses different masses for 
particles and antiparticles the theory must be
nonlocal in terms of the $x$-space fields associated with the 
particles.
In that case the propagator is not covariant, and, further, the lack
of covariance leads to qualitatively different behaviors of the 
propagator at large $s$ in different frames of reference.

We are happy to thank Rabi Mohapatra for suggesting that we consider 
the relation between $CPT$ and Lorentz invariance. We
acknowledge helpful discussions with E. Boldt, T. Cohen,
P. Frampton, T. Jacobson, W. Kummer, R. Mohapatra, S. Nussinov and 
A. Shimony.


\begin{thebibliography}{20}
\bibitem{cone} By ``in cone'' we mean that the momenta that are
conjugate to the successive spacetime difference variables are
physical, i.e., are in or on the forward light cone in momentum
space. Similarly, ``out of cone'' means that unphysical momenta that
can be outside the forward light cone are present in addition 
(possibly) to physical momenta.
\bibitem{particle} If we consider the theta functions as external fields,
then the propagator is invariant under particle Lorentz transformations.
\end{thebibliography}
\end{document}